\documentclass[12pt,a4paper,showpacs]{revtex4}
\usepackage{amsmath}
\usepackage{amssymb}
\usepackage{bm}
\usepackage{graphicx}
\begin{document}

\title{%
Computer Simulation of the Critical Behavior
of 3D Disordered Ising Model
}

\author{
Vladimir V. \textsc{Prudnikov}\footnote{E-mail: prudnikv@univer.omsk.su},
Pavel V. \textsc{Prudnikov},
Andrei N. \textsc{Vakilov} and
Alexandr S. \textsc{Krinitsyn}%
}

\affiliation{%
Dept. of Theoretical Physics, Omsk State University, Omsk 644077, Russia
}

\begin{abstract}
The critical behavior of the disordered ferromagnetic Ising model is studied numerically by the
Monte Carlo method in a wide range of variation of concentration of nonmagnetic impurity atoms. The temperature
dependences of correlation length and magnetic susceptibility are determined for samples with various
spin concentrations and various linear sizes. The finite-size scaling technique is used for obtaining scaling functions
for these quantities, which exhibit a universal behavior in the critical region; the critical temperatures and
static critical exponents are also determined using scaling corrections. On the basis of variation of the scaling
functions and values of critical exponents upon a change in the concentration, the conclusion is drawn concerning
the existence of two universal classes of the critical behavior of the diluted Ising model with different characteristics
for weakly and strongly disordered systems.
\end{abstract}


\maketitle

%%%%%%%%%%%%%%%%%%%%%%%%%%%%%%%%%%%%%%%%%%%%%%%%%%%%%%%%%%%%%%%%%%%%%%%%%%%%%%%%%%%%%%%%%%%%%%%%%%%%%%%%%%%%%%%%%%%%%%%%%
\section{Introduction}
%%%%%%%%%%%%%%%%%%%%%%%%%%%%%%%%%%%%%%%%%%%%%%%%%%%%%%%%%%%%%%%%%%%%%%%%%%%%%%%%%%%%%%%%%%%%%%%%%%%%%%%%%%%%%%%%%%%%%%%%%

Analysis of the critical behavior of disordered systems
with quenched structural defects is of considerable
theoretical and experimental interest. Most real
solids contain quenched structural defects, whose presence
can affect the characteristics of the system and
may strongly modify the behavior of the systems during
phase transitions. This leads to new complex phenomena
in structurally disordered systems, which are
associated with the effects of an anomalously strong
interaction of fluctuations of a number of thermodynamic
quantities, when any perturbation introduced by
structural defects (even in small concentration) may
strongly change the state of the system. The description
of such systems requires the development of special
analytic and numerical methods.

The following two questions arise when the effect of
structural disorder on second-order phase transitions is
investigated: (i) do the critical exponents of a homogeneous
magnet change upon its dilution by nonmagnetic
impurity atoms? and (ii) if these exponents change, are
the new critical exponents universal (i.e., independent
of the structural defect concentration up to the percolation
threshold)? The answer to the first question was
obtained in \cite{Harris}, where it was shown that the critical
exponents of systems with quenched structural defects
change as compared to their homogeneous analogs if
the critical exponent of the heat capacity of a homogeneous
system is positive. This criterion is satisfied only
by 3D systems whose critical behavior can be described
by the Ising model. A large number of publications are
devoted to the study of the critical behavior of diluted
Ising-like magnets by the renorm-group methods, the
numerical Monte Carlo methods, and experimentally
(see review \cite{Holovatch}). An affirmative answer has been
obtained to the question concerning the existence of a
new universal class of the critical behavior, which is
formed by diluted Ising-like magnet. It remains
unclear, however, whether the asymptotic values of
critical exponents are independent of the rate of dilution
of the system, how the crossover effects change
these values, and whether two or more regimes of the
critical behavior exist for weakly and strongly disordered
systems; these questions are the subjects of
heated discussions.

This study is devoted to numerical analysis of the
critical behavior of a diluted 3D Ising model in a wide
range of concentration of quenched point defects. The
fundamental importance of the results of this study is
due to stringent requirements to simulation conditions
imposed in the course of investigations; the wide range
of linear dimensions of lattices ($L=20 - 400$) analyzed
in this work; the chosen temperature range of simulation
close to the critical temperature with
$\tau=(T-T_c)/T_c = 5\cdot 10^{-4} - 10^{-2}$, which makes it possible to single out the
asymptotic values of characteristics; high statistics
used for averaging of thermodynamic and correlation
functions over various impurity configurations; the
application of finite-size scaling technique \cite{Landau} for processing
the result of simulation, which makes it possible
to obtain scaling function for thermodynamic functions
apart from their asymptotic values; and application
of corrections to scaling for determining the
asymptotic values of critical exponents.

%%%%%%%%%%%%%%%%%%%%%%%%%%%%%%%%%%%%%%%%%%%%%%%%%%%%%%%%%%%%%%%%%%%%%%%%%%%%%%%%%%%%%%%%%%%%%%%%%%%%%%%%%%%%%%%%%%%%%%%%%
\section{Computer simulation technique and results}
%%%%%%%%%%%%%%%%%%%%%%%%%%%%%%%%%%%%%%%%%%%%%%%%%%%%%%%%%%%%%%%%%%%%%%%%%%%%%%%%%%%%%%%%%%%%%%%%%%%%%%%%%%%%%%%%%%%%%%%%%

We consider a model of a disordered spin system in
the form of a cubic lattice with linear size
$L$ under certain boundary conditions. The microscopic Hamiltonian
of the disordered Ising model can be written in the
form
\begin{equation}
H=\frac{1}{2}\sum\limits_{i,j}J_{ij}\sigma_{i}\sigma_{j}p_{i}p_{j},
\label{f_H}
\end{equation}
where
$J_{ij}$ is the short-range exchange interaction
between spins $\sigma_{i}$
fixed at the lattice sites and assuming
values of $\pm 1$. Nonmagnetic impurity atoms form empty
sites. In this case, occupation numbers
$p_{i}$ assume the
value $0$ or $1$ and are described by the distribution function
\begin{equation}
P(p_{i})=(1-p)\delta(p_{i})+p\delta(1-p_{i})
\label{f_P}
\end{equation}
with $p=1-c$, where $c$ -
is the concentration of the impurity
atom. The impurity is uniformly distributed over
the entire system, and its position is fixed in simulation
for an individual impurity configuration. We consider
here disordered systems with spin concentrations $p=0.95$, $0.80$, $0.60$, and $0.50$.

To suppress the effect of critical slowing down and
correlation of various spin configurations, we used the
single-cluster Wolf algorithm, which is most effective
in this respect \cite{Hennecke,Ivaneyko}. A Monte Carlo step per spin
(MCS) was assumed to correspond to 10–20 rotations
of a Wolf cluster depending on the linear size of the lattice
being simulated, the spin concentration of the system,
and the closeness of the temperature to the critical
point. The stabilization of thermodynamic equilibrium
required $10^{4}$ Monte Carlo steps, and  $10^{5}$ steps were
allotted to statistical averaging of quantities being simulated
for a given impurity configuration. To determine
the average values of thermodynamic and correlation
functions, averaging over various impurity configurations
was carried out along with statistical averaging
(averaging was carried out over $3000$ samples for
$p=0.95$, over $5000$ samples for
$p=0.80$, and over $10000$ samples for $p=0.60$ and $0.50$).

In the course of simulation of various spin systems,
correlation length $\xi_L$
and susceptibility $\chi_L$
were carried out on lattices with a linear size $L$ in accordance with
the following relations:
\begin{equation}
\xi=\frac{1}{2\sin(\pi/L)}\sqrt{\frac{\chi}{F}-1},\qquad\chi=\frac{1}{pL^{3}}\overline{<S^{2}>},
\label{f_xi}
\end{equation}
where $S=\sum\limits_{i}p_{i}\sigma_{i}$, $F=\overline{<\Phi>}/pL^{3}$, and
\begin{equation}
\Phi=\frac{1}{3}\sum\limits_{n=1}^{3}\left|\sum\limits_{i}p_{i}\sigma_{i}\exp\left(\frac{2\pi i x_{n,i}}{L}\right)\right|^{2},
\label{f_Phi}
\end{equation}
where $(x_{1,i},x_{2,i},x_{3,i})$ are the coordinates of the
$i$-th lattice site; angle brackets indicate statistical averaging
over Monte Carlo steps, and the bar indicates averaging
over impurity configurations. The temperature dependences
$\xi_L(T)$ and $\chi_L(T)$ were determined in the temperature
interval  $\tau= 5\cdot 10^{-4} - 10^{-2}$
for samples with $p=0.95$ and a linear size in the range of
$L=20-400$. For
samples with the remaining spin concentrations, temperatures
were chosen in the interval of $\tau=10^{-3} - 10^{-2}$
for values of $L$ ranging from $20$ to $300$. In computer
simulation, the value of $L_{max}$ for each temperature was
limited by the lattice size for which the correlation
length and susceptibility of the system attained their
asymptotic values.

In accordance with the results obtained in \cite{Hennecke,Ivaneyko} and
the results of our investigations, the chosen simulation
conditions ensure equilibrium values for measurable
thermodynamic quantities for all lattice sizes and spin
concentrations studied here since the autocorrelation
times for magnetization and energy turn out to be not
longer than ten Monte Carlo steps per spin even for
chosen temperatures closest to the critical temperature
(with allowance for the number of turns of the Wolf
cluster taken as a step).

%%%%%%%%%%%%%%%%%%%%%%%%%%%%%%%%%%%%%%%%%%%%%%%%%%%%%%%%%%%%%%%%%%%%%%%%%%%%%%%%%%%%%%%%%%%%%%%%%%%%%%%%%%%%%%%%%%%%%%%%%
\section{METHOD OF FINITE-SIZE SCALING}
%%%%%%%%%%%%%%%%%%%%%%%%%%%%%%%%%%%%%%%%%%%%%%%%%%%%%%%%%%%%%%%%%%%%%%%%%%%%%%%%%%%%%%%%%%%%%%%%%%%%%%%%%%%%%%%%%%%%%%%%%
It is known that the second-order phase transition
considered here can be manifested only in the thermodynamic
limit, when the volume of the system and the
number of particles in it tend to infinity. To determine
the asymptotic values of thermodynamic quantities
$A(T)$ exhibiting an anomalous behavior near the critical
temperature from their values $A_{L}(T)$ determined on
finite lattices, the concepts of the scaling theory concerning
the generalized uniformity of thermodynamic
functions in the critical region relative to scale transformations
of the system are widely used. These concepts
formed the basis of various methods of finite-size scaling.
Here, we apply the method proposed in \cite{Landau} and
tested by the authors in analysis of the results of simulation
of the critical behavior of 2D and 3D pure Ising
models.

The idea of this method \cite{Landau} is that, in accordance
with the scaling theory, the size dependence of a certain
thermodynamic quantity $A_{L}$ defined on a finite lattice in
zero magnetic field can be presented in the critical
region in the form
\begin{equation}
A_{L}(\tau)=L^{\delta/\nu}f_{A}(s_{L}(\tau)),\qquad s_{L}(\tau)=L/\xi_L(\tau),
\label{f_A_L_s}
\end{equation}
where $\delta$ is the critical exponent for the thermodynamic
quantity $A(\tau)\sim\tau^{-\delta}$.
Taking into account the fact that the
correlation length in the critical region behaves as
$\xi(\tau)\sim\tau^{-\nu}$, we can write
\begin{equation}
L^{\delta/\nu}=A(\tau)s_{L}^{\delta/\nu}(\tau).
\label{f_L}
\end{equation}

\begin{figure}[t]
\includegraphics[width=0.8\textwidth]{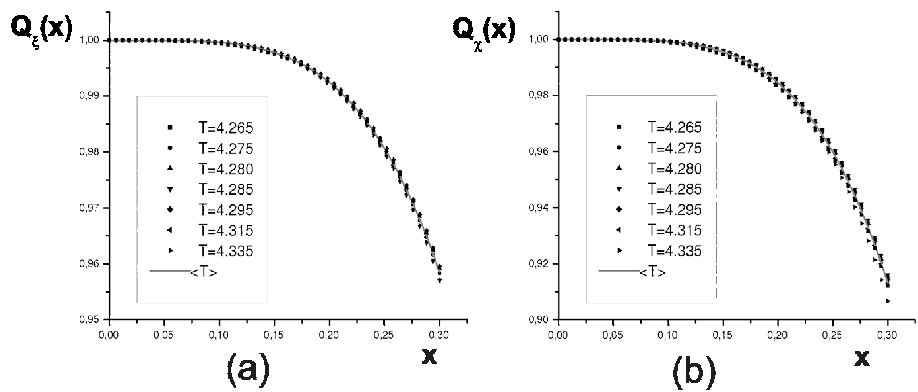}
\caption{\label{fig:1}Scaling functions for (a) correlation length and (b) susceptibility obtained at various temperatures for a system with $p=0.95$
using the approximation polynomial in $x$.}
\end{figure}

Then expression (\ref{f_A_L_s}) can be written in the form
\begin{equation}
A_{L}(\tau)=A(\tau)F_{A}(s_{L}(\tau)),
\label{f_A_L_F}
\end{equation}
where the relation between scaling functions $f_{A}$ and $F_{A}$ is defined in the form of the relation
\begin{equation}
F_{A}(s_{L}(\tau))=s_{L}^{\delta/\nu}(\tau)f_{A}(s_{L}(\tau)).
\label{f_F_A}
\end{equation}
If correlation length $\xi$ plays the role of quantity $A$,
Eq.~(\ref{f_A_L_F}) defines $\xi_{L}(\tau)/L$ as a function of only one variable $\xi(\tau)/L$.
This leads to a relation that makes it possible
to find the asymptotic value of any thermodynamic
quantity in terms of directly measurable values of $A_{L}$ and the scaling function of $x_{L}(\tau)=\xi_{L}(\tau)/L$,
\begin{equation}
A(\tau)=A_{L}(\tau)/Q_{A}(x_{L}(\tau)),
\label{f_A_L_Q}
\end{equation}
where function $Q_{A}(x_{L}(\tau))$ is defined by the expression
\begin{equation}
Q_{A}(x_{L}(\tau))=F_{A}(f_{\xi}^{-1}(x_{L}(\tau))).
\label{f_Q_A}
\end{equation}
Scaling function $Q_{A}(x_{L})$, defined in the interval $0\leq x_{L} \leq x_c$, where
$x_c$ is the value of the argument independent of $L$ in
the critical region, must satisfy the following asymptotic
conditions:
$\lim\limits_{x\to 0}Q_A(x) \to 1$ and $\lim\limits_{x \to x_c}Q_A \to 0$.

\begin{figure}[t]
\includegraphics[width=0.8\textwidth]{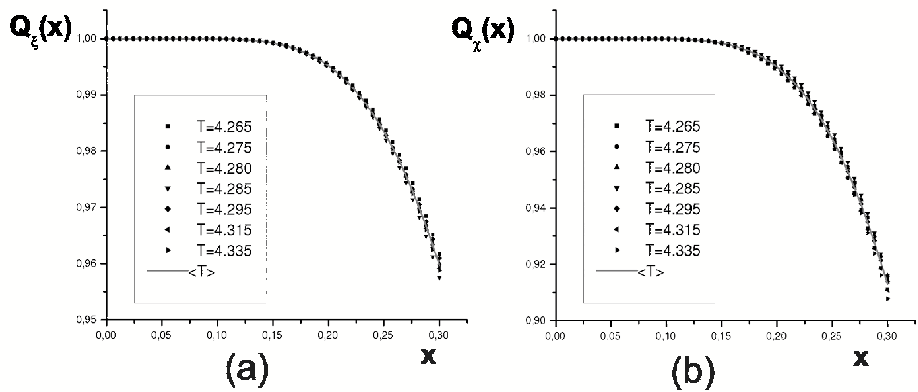}
\caption{\label{fig:2}Scaling functions for (a) correlation length and (b) susceptibility obtained at various temperatures for a system with $p=0.95$
using the approximation polynomial in $\exp(-1/x)$.}
\end{figure}

To satisfy the asymptotic conditions, we chose,
analogously to \cite{Landau}, the scaling function for susceptibility
and correlation length in the form of a polynomial
dependence of $x$, as well as of
$\exp\left(-1/x\right)$ :
\begin{eqnarray}
Q_A(x)&=&1+c_{1}x+c_{2}x^2+c_{3}x^3+c_{4}x^4, \\
Q_A(x)&=&1+c_{1}e^{-1/x}+c_{2}e^{-2/x}+c_{3}e^{-3/x}+c_{4}e^{-4/x},
\end{eqnarray}
with coefficients $c_n$ selected for each temperature $T$ using the least squares method.

\begin{figure}[t]
\includegraphics[width=0.8\textwidth]{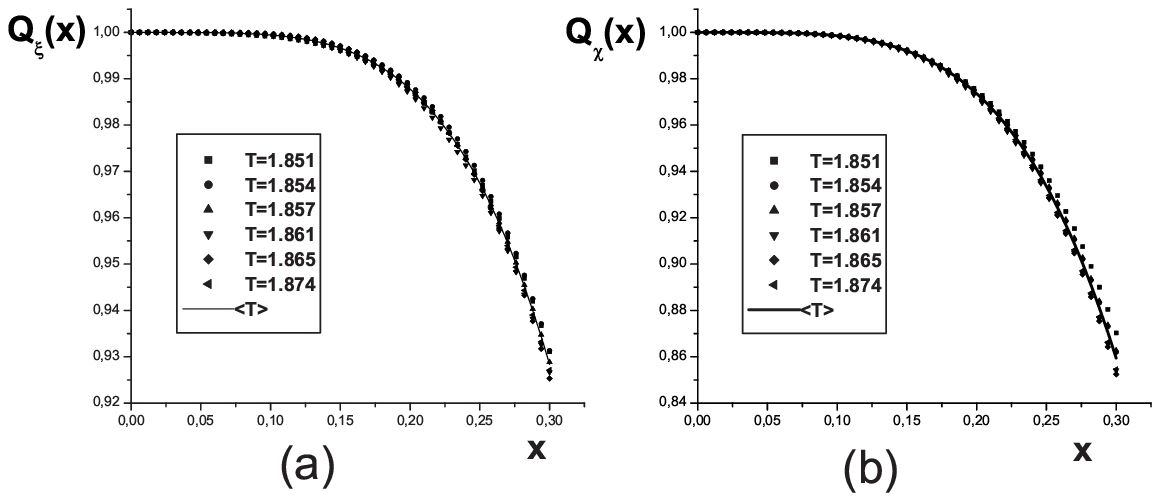}
\caption{\label{fig:3}Scaling functions for (a) correlation length and (b) susceptibility obtained at various temperatures for a system with $p=0.50$
using an approximation polynomial in $x$.}
\includegraphics[width=0.8\textwidth]{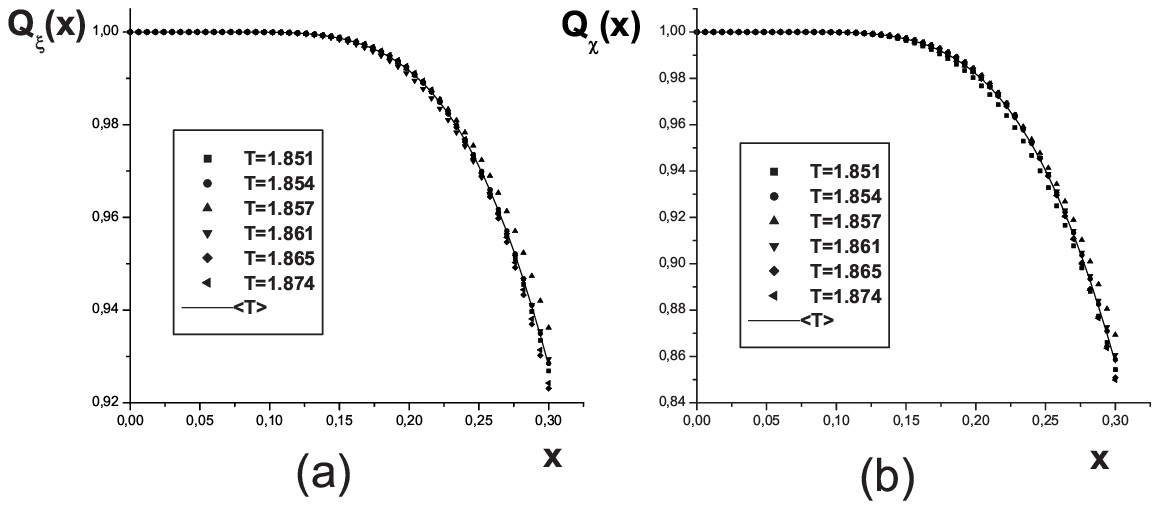}
\caption{\label{fig:4}Scaling functions for (a) correlation length and (b) susceptibility obtained at various temperatures for a system with $p=0.50$
using the approximation polynomial in $\exp(–1/x)$.}
\end{figure}

Here, we implement the following scheme of finitesize
scaling.
\begin{enumerate}
\item For an arbitrary value of $\tau_0$ in the critical temperature range, the values of
$A_L(\tau_0)$ and $x(L,\tau_0)=\xi_L(\tau_0)/L$ are measured for lattices with increasing size $L$.
\item The thermodynamic value of quantity $A(\tau_0)$ is determined as the value of $A_L(\tau_0)$,
which is found to be independent of $L$ within the error of measurements.
\item The results of measurements for $A_L(\tau_0)/A(\tau_0)$
are processed by the least squares method to determine the
corresponding functional form for scaling function $Q_A(x(L,\tau_0)$.
\item The procedure is repeated for other values of $\tau$ in the range of $\tau \simeq 10^{-3} - 10^{-2}$.
\item Averaged scaling function $Q_{\text{aver}}^A$ is determined
on the basis of functions $Q_A(x(L,\tau_i)$, determined for
various temperatures $\tau_i$ for a fixed spin concentration $p$ of the samples.
\item The temperature dependence is determined for
asymptotic values of the thermodynamic quantity by
substituting $A_L(\tau)$ and $Q^A_{\text{aver}}$ into relation (\ref{f_A_L_Q}).
\end{enumerate}

\begin{figure}[t]
\includegraphics[width=0.45\textwidth]{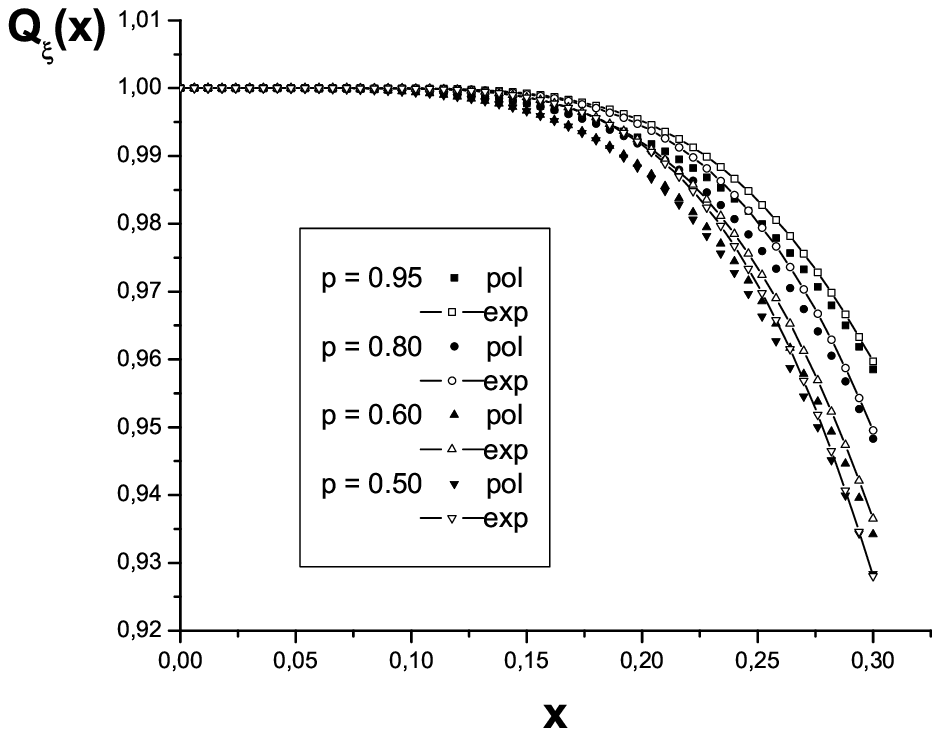}
\quad
\includegraphics[width=0.45\textwidth]{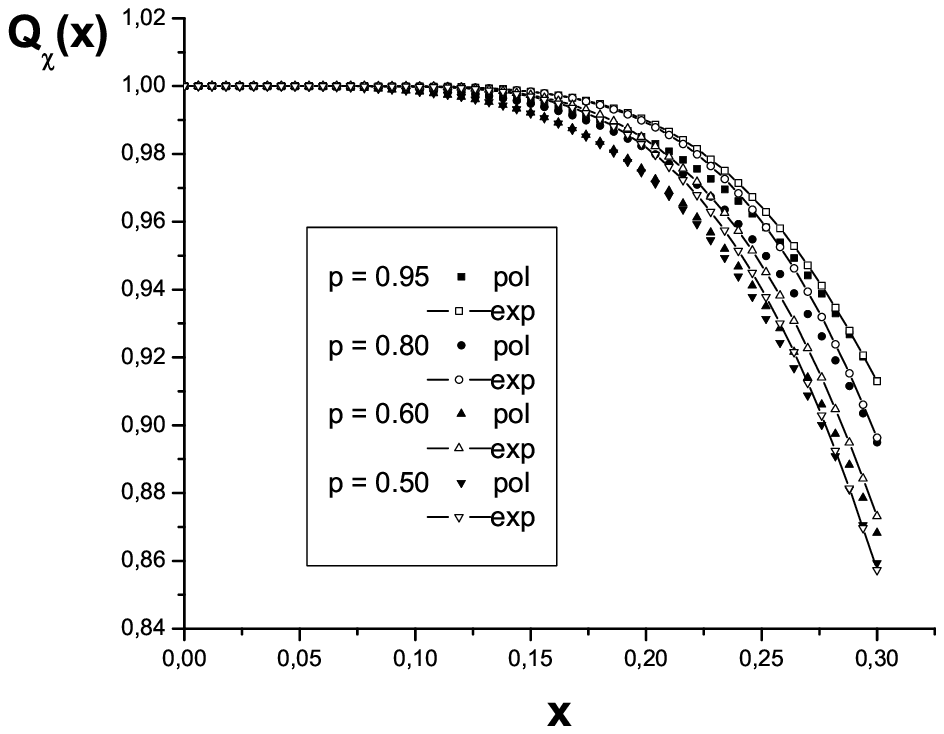}
\parbox{0.45\textwidth}{
\caption{\label{fig:5}Averaged scaling functions for correlation length
obtained using the approximation polynomial in ($x$) (symbols)
and in $\exp(–1/x)$ solid curves.}}
\quad
\parbox{0.45\textwidth}{
\caption{\label{fig:6}Averaged scaling functions for susceptibility
obtained using the approximation polynomial in ($x$) (symbols)
and in $\exp(–1/x)$ solid curves.}}
\end{figure}

Figures~\ref{fig:1}-\ref{fig:4} show by way of example the scaling
functions for correlation length $\xi$ and susceptibility $\chi$,
obtained for systems with spin concentrations $p=0.95$
and $0.50$ at various temperatures using the polynomial
approximation in variable $x$ (Figs.~\ref{fig:1}, \ref{fig:3}) and in variable $\exp\left(-1/x\right)$ (Figs.~\ref{fig:2}, \ref{fig:4}).
It can be seen from the figures
that the scaling functions show a tendency towards a
single universal curve for each spin concentration $p$
in the entire range of variation of scaling variable $x_L$.

Figures 5 and 6 show the averaged scaling functions for the
correlation length and susceptibility for various spin
concentrations $p$, which were obtained using the polynomial
approximation in $x$ (Fig.~\ref{fig:5}) and in $\exp\left(-1/x\right)$ (Fig.~\ref{fig:6}).
The averaged scaling functions demonstrate a
tendency indicating the possible existence of two
classes of universal critical behavior for the diluted
Ising model with different modes of behavior for
weakly ($p=0.95, 0.80$) and strongly ($p = 0.60, 0.50$)
disordered systems.

Table~\ref{tab:1} contains asymptotic values of $\xi(T)$ and $\chi(T)$
obtained using averaged scaling functions for various
temperatures and spin concentrations. The errors in values
of $\xi(T)$ and $\chi(T)$ take into account statistical errors
in the measured values of $\xi_L(T)$ and $\chi_L(T)$, as well as
approximation errors.

\begin{table}[t]
  \caption{\label{tab:1}Asymptotic values of correlation length and susceptibility obtained using scaling functions with a polynomial
dependence on $x$ (pol) and $\exp(-1/x)$ (exp).}
  \small
    \begin{tabular}{cc|lllllll}
      \hline\hline
      p=0.95& T &  4.265  &  4.275  &  4.280  &  4.285  &  4.295 &  4.315 & 4.335 \\
      \hline\hline
      $ \xi$&pol&62.44(15)& 21.25(5)& 17.06(4)& 14.41(3)&11.40(2)& 8.37(2)&6.76(2)\\
            &exp&62.31(15)& 21.19(6)& 17.02(4)& 14.38(3)&11.38(2)& 8.36(2)&6.76(2)\\
      $\chi$&pol&14467(80)& 1748(16)&  1130(4)& \ 819(3)&\ 515(2)&  282(2)&187(1) \\
            &exp&14359(81)& 1724(12)&  1126(5)& \ 813(4)&\ 513(2)&  281(2)&187(1) \\
      \hline\hline
      p=0.80& T &  3.51   &  3.52   &  3.53   &  3.54   &  3.55  &  3.57  &       \\
      \hline\hline
      $ \xi$&pol& 26.16(9)& 16.50(4)& 12.51(2)& 10.31(2)& 8.79(3)& 7.01(3)&       \\
            &exp& 26.11(9)& 16.46(4)& 12.49(3)& 10.30(2)& 8.76(3)& 7.00(3)&       \\
      $\chi$&pol& 2612(17)&  1060(5)& \ 618(2)& \ 424(2)&  312(2)&  201(2)&       \\
            &exp& 2603(18)&  1055(5)& \ 615(2)& \ 423(2)&  310(2)&  200(1)&       \\
      \hline\hline
      p=0.60& T &  2.430  &  2.435  &  2.440  &  2.445  &  2.450 &  2.460 &       \\
      \hline\hline
      $ \xi$&pol&46.03(15)& 29.37(7)& 22.49(8)&18.33(5) &15.70(4)&12.41(4)&       \\
            &exp&45.86(13)& 29.29(7)& 22.40(8)&18.27(5) &15.65(5)&12.37(4)&       \\
      $\chi$&pol& 7943(55)& 3289(17)& 1953(15)& 1308(7) &\ 967(5)&\ 611(4)&       \\
            &exp& 7881(47)& 3268(15)& 1937(14)& 1298(7) &\ 961(5)&\ 608(4)&       \\
      \hline\hline
      p=0.50& T &  1.851  &  1.854  &  1.857  &  1.861  &  1.865 &  1.874 &       \\
      \hline\hline
      $ \xi$&pol&49.55(46)&36.38(12)&29.57(9) & 23.86(6)&20.23(4)&15.38(3)&       \\
            &exp&49.04(38)&36.34(16)&29.48(11)& 23.81(7)&20.17(4)&15.35(3)&       \\
      $\chi$&pol&9456(356)& 5036(36)& 3365(26)& 2209(12)& 1603(6)&\ 939(4)&       \\
            &exp&9169(165)& 5030(43)& 3349(25)& 2199(12)& 1592(6)&\ 934(3)&       \\ \hline\hline
    \end{tabular}
\end{table}

%%%%%%%%%%%%%%%%%%%%%%%%%%%%%%%%%%%%%%%%%%%%%%%%%%%%%%%%%%%%%%%%%%%%%%%%%%%%%%%%%%%%%%%%%%%%%%%%%%%%%%%%%%%%%%%%%%%%%%%%%
\section{CALCULATION OF CRITICAL CHARACTERISTICS}
%%%%%%%%%%%%%%%%%%%%%%%%%%%%%%%%%%%%%%%%%%%%%%%%%%%%%%%%%%%%%%%%%%%%%%%%%%%%%%%%%%%%%%%%%%%%%%%%%%%%%%%%%%%%%%%%%%%%%%%%%

Asymptotic critical exponent of a thermodynamic
quantity $A(\tau)$ is described by the expression
\begin{equation}
\delta=-\lim\limits_{\tau\rightarrow 0}\frac{\ln A(\tau)}{\ln|\tau|},\qquad A(\tau)=A_{\pm}|\tau|^{-\delta},
\label{f_x}
\end{equation}
where $A_{+}$ and $A_{-}$ are the critical amplitudes above and
below the critical temperature, respectively. A power
law of the type (\ref{f_x}) is accurate only in the limit $\tau\rightarrow 0$.
To calculate critical exponents in the intermediate
nonasymptotic regime, we must introduce additional
correcting terms to power law (\ref{f_x}).
In accordance with
the Wegner expansion \cite{Wegner} we have
\begin{equation}
A(\tau)=\left(A_{0}+A_{1}\tau^{\omega\nu}+A_{2}\tau^{2\omega\nu}+\dots\right)\tau^{-\delta}\ (\tau>0),
\label{f_A_weg}
\end{equation}
where $A_{i}$ are nonuniversal amplitudes and $\omega$ is the critical
exponent of the correction to scaling. Here, in the
calculation of the characteristics of disordered systems,
we use the first correction to the asymptotic behavior
for the correlation length and susceptibility:
\begin{equation}
\xi(\tau)=\tau^{-\nu} \left(A_{0}^{\xi}+A_{1}^{\xi}\tau^{\theta}\right),\qquad\theta=\omega\nu,
\label{f_Axi}
\end{equation}
\begin{equation}
\chi(\tau)=\tau^{-\gamma} \left(A_{0}^{\chi}+A_{1}^{\chi}\tau^{\theta}\right),\phantom{\qquad\theta=\omega\nu,}
\label{f_Achi}
\end{equation}
and calculated critical exponents $\nu$, $\gamma$ and $\theta$,
as well as
the critical temperatures using the least squares method
for the best approximation of the data presented in
Table~\ref{tab:1} by expressions (\ref{f_Axi}) and (\ref{f_Achi}).
Table~\ref{tab:2} contains
the values of critical parameters obtained for various
spin concentrations $p$ using the initial data corresponding
to various approximations for scaling functions, as
well as their values averaged over approximations. It
can be seen that the critical exponents form two groups
with close values to within experimental error. The first
group corresponds to $p=0.95$ and $0.80$, i.e., to weakly
disordered systems with spin concentrations $p$, larger
than the impurity percolation threshold $p_{imp}$ ($p_{imp}=0.69$ for cubic systems),
while the second group with $p=0.60$ and $0.50$, corresponds to strongly disordered systems
with $p_c < p < p_{imp}$, where $p_c$ is the spin percolation
threshold ($p_c=0.31$ for cubic systems), ; in the latter
case, two mutually penetrating (spin and impurity)
clusters exist in the system. Fractal effects of these two
penetrating clusters may be responsible for the change
in the type of critical behavior for strongly disordered
systems. We can consider that the averaged values of
critical exponents $\nu=0.693(5)$, $\gamma=1.342(7)$ and $\theta=0.157(92)$ for weakly disordered systems and
$\nu=0.731(11)$, $\gamma=1.422(12)$ and $\theta=0.203(106)$ for
strongly disordered systems are the final results of our
investigations.
It should be noted that the values of the
critical exponents obtained for weakly disordered systems
correlate with the values of $\nu=0.678(10)$, $\gamma=1.330(17)$ and
$\theta=0.170(71)$ ($\omega=0.25(10)$), obtained in \cite{Pelissetto} by the renormalizations group methods in the six-loop
approximation, which are valid only for systems with
low concentrations of impurities.

\begin{table}[p]
\centering
  \caption{\label{tab:2}Values of critical parameters for two types of approximations (pol) and (exp) and their averaged (aver) values for
systems with various spin concentrations $p$}
  \small
    \begin{tabular}{cc|llllll}
      \hline\hline
            &    &$   \nu  $&$  \gamma $&$\theta^{\xi}$&$\theta^{\chi}$&$T_{c}^{\xi}$&$T_{c}^{\chi}$\\
      \hline\hline
      p=0.95& pol&0.6883(15)&1.3339(25) &   0.141(52)  &   0.152(50)   & 4.26264(4)  &  4.26269(3)  \\
            & exp&0.6935(26)&1.3430(33) &   0.113(64)  &   0.142(54)   & 4.26265(5)  &  4.26270(3)  \\
            &aver&0.6909(33)&1.3385(54) &   0.137(56)  &               & 4.26267(4)  &              \\
      \hline\hline
      p=0.80& pol&0.6960(29)&1.3473(30) &   0.180(107) &   0.193(74)   & 3.49937(21) &  3.49954(14) \\
            & exp&0.6947(28)&1.3421(30) &   0.147(94)  &   0.192(71)   & 3.49940(21) &  3.49961(14) \\
            &aver&0.6956(29)&1.3447(40) &   0.178(87)  &               & 3.49948(18) &              \\
      \hline\hline
      p=0.60& pol&0.7272(37)&1.4253(34) &   0.221(147) &   0.201(63)   & 2.42409(11) &  2.42404(6)  \\
            & exp&0.7233(24)&1.4054(43) &   0.184(92)  &   0.192(109)  & 2.42414(8)  &  2.42423(7)  \\
            &aver&0.7253(36)&1.4154(107)&   0.199(103) &               & 2.42413(9)  &              \\
      \hline\hline
      p=0.50& pol&0.7372(25)&1.4299(26) &   0.164(159) &   0.195(74)   & 1.84503(7)  &  1.84512(3)  \\
            & exp&0.7368(26)&1.4266(30) &   0.242(96)  &   0.226(66)   & 1.84503(7)  &  1.84519(3)  \\
            &aver&0.7370(33)&1.4283(33) &   0.207(100) &               & 1.84509(6)  &              \\
      \hline\hline
    \end{tabular}
\vspace*{5mm}  \caption{\label{tab:3}Values of critical exponents $\nu$ and $\gamma$ experimentally measured by different authors in materials corresponding to
the disordered Ising model}
  \small
    \begin{tabular}{c|cccc}
      \hline\hline
                Authors          &$Fe_{p}Zn_{1-p}F_{2}$&$             |\tau|             $&$     \nu    $&$   \gamma   $\\
      \hline\hline
      Birgeneau et al.,         &        p=0.60       &$       10^{-1}-2   \cdot 10^{-3}$&$0.73(3)$&$1.44(6)$\\
      1983, \cite{Birgeneau}    &        p=0.50         &$2\cdot 10^{-2}-2   \cdot 10^{-3}$&              &              \\
      Belanger et al.,          &        p=0.46       &$       10^{-1}-1.5 \cdot 10^{-3}$&$0.69(1)$&$1.31(3)$\\
      1986, \cite{Belanger_1986}&                     &                                  &              &              \\
      Slanic et al.,            &        p=0.93       &                                  &$0.71(1)$&$1.35(1)$\\
      1998, \cite{Belanger_1998}&                     &                                  &              &              \\
      Slanic et al.,            &        p=0.93       &$       10^{-2}-1.14\cdot 10^{-4}$&$0.70(2)$&$1.34(6)$\\
      1999, \cite{Slanic}       &                     &                                  &              &              \\
      \hline\hline
                                &$Mn_{p}Zn_{1-p}F_{2}$&$            |\tau|           $&$      \nu     $&$    \gamma    $\\
      \hline\hline
      Mitchell et al.,     &        p=0.75       &$2\cdot 10^{-1}-4\cdot 10^{-4}$&$0.715(35)$&$1.364(76)$\\
      1986, \cite{Mitchell}&        p=0.50       &$1\cdot 10^{-1}-5\cdot 10^{-3}$&$0.75(5) $&$1.57(16) $\\
      \hline\hline
    \end{tabular}
\end{table}

The above values of critical exponents $\nu$ and $\gamma$ are
also in good agreement with the available results of
experiments with diluted Ising-like magnets (Table~\ref{tab:3}).

%%%%%%%%%%%%%%%%%%%%%%%%%%%%%%
\begin{table}[t]
\centering
  \caption{\label{tab:4} Values of critical exponents $\nu$ and $\gamma$ obtained by different authors using the Monte Carlo method}
  \small
    \begin{tabular}{l|ccccc}
      \hline\hline
         \multicolumn{1}{c|}{Authors}         &$L_{max}$&$        p        $&$       \nu      $&$    \gamma    $& $\theta=\omega\nu$\\
      \hline\hline
                Wang et al., 1990, \cite{Wang}&   300   &$       0.8       $&                  &$1.36(4) $ &\\
                      Heuer, 1993,\cite{Heuer}&   60    &      $0.95$       &     $0.64(2)$    &$1.28(3)$  &    \\
                                              &         &      $0.9$        &     $0.65(2)$    &$1.31(3)$  &   \\
                                              &         &      $0.8$        &     $0.68(2)$    &$1.35(3)$  &  \\
                                              &         &      $0.6$        &     $0.72(2)$    &$1.51(3)$  & \\
  Wiseman et al.,1998, \cite{Wiseman}         &    64   &$       0.8       $&     $0.682(3)$   &$1.357(8)$ & \\
                                              &    80   &$       0.6       $&     $0.717(7)$   &$1.508(28)$&   \\
 Ballesteros et al., 1998, \cite{Ballesteros} &   128   &$0.4\leq p\leq 0.9$&     $0.6837(53)$ &$1.342(10)$& $0.253(43)$\\
     Calabrese et al., 2003, \cite{Calabrese} &    256  &        0.8        &$0.683(3)        $&$1.336(8)$ & $0.581(85)$ \\
Berche et al., 2005, \cite{Ivaneyko5}         &     96  &        0.85       &$0.662(2)        $&$1.314(4)$ & \\
Murtazaev et al., 2004, \cite{Murtazaev}       &   60    &      $0.95$       &     $0.646(2)$   &$1.262(2)$ &      \\
                                              &         &      $0.9$        &     $0.664(2)$   &$1.285(3)$ &      \\
                                              &         &      $0.8$        &     $0.683(4)$   &$1.299(3)$ &      \\
                                              &         &      $0.6$        &     $0.725(6)$   &$1.446(4)$ &      \\
      \hline\hline
    \end{tabular}
\end{table}
%%%%%%%%%%%%%%%%%%%%%%%%%%%%%%

Table~\ref{tab:4} contains the latest results of Monte Carlo
simulation of the critical behavior of the diluted Ising
model obtained by various authors. Each work cited in
the table has its merits due to the application of various
techniques for processing the results of simulation, as
well as disadvantages associated with the small size of
experimental lattices, which does not ensure reliable
asymptotic values of the quantities being measured, or
with insufficient statistics of averaging over various
impurity configurations for obtaining reliable results,
or with disregard of the effect of nonasymptotic corrections
to scaling in the calculation of critical exponents
(the inclusion of these corrections is especially important
for samples with spin concentrations
$p=0.95$ and $0.90$ and strongly disordered systems).
The results
obtained in \cite{Heuer,Murtazaev}
can be treated as supporting our
conclusions; these ideas were formulated in our earlier
publications \cite{Prudnikov} on computer simulation of critical
dynamics of the disordered Ising model. In fact, in spite
of the attractiveness of the idea about a single universal
critical behavior with the asymptotic values of critical
exponents independent of the spin concentration,
which was supported by the authors of \cite{Ballesteros}, the results
obtained in \cite{Ballesteros} did not make it possible to adequately
explain the results obtained for samples with $p=0.90$ using the universal
critical exponent of scaling correction $\omega=0.37(6)$ for all systems.
At the same time, the nonasymptotic values of critical exponents obtained in \cite{Ballesteros}
demonstrated explicit dependence on $p$ and,
assuming that $\omega$ is not unique, led to two sets of asymptotic
critical exponents for weakly and strongly disordered
systems. The results of the remaining studies carried
out on samples with a single spin concentration
coincide as a rule with our results to within experimental
error, although some mismatching associated in all
probability with the above-mentioned drawbacks also
takes place.

%%%%%%%%%%%%%%%%%%%%%%%%%%%%%%%%%%%%%%%%%%%%%%%%%%%%%%%%%%%%%%%%%%%%%%%%%%%%%%%%%%%%%%%%%%%%%%%%%%%%%%%%%%%%%%%%%%%%%%%%%
\section{CONCLUSIONS}
%%%%%%%%%%%%%%%%%%%%%%%%%%%%%%%%%%%%%%%%%%%%%%%%%%%%%%%%%%%%%%%%%%%%%%%%%%%%%%%%%%%%%%%%%%%%%%%%%%%%%%%%%%%%%%%%%%%%%%%%%

The results of our investigations lead to the following
conclusions.
\begin{description}
\item[(i)]
Scaling functions and values of critical exponents
for the correlation length and susceptibility demonstrate
the existence of two classes of universal critical behavior
for the diluted Ising model with various
characteristics for weakly and strongly disordered systems.
\item[(ii)] The values of critical exponents obtained for
weakly disordered systems are in good agreement with
the results of the field-theoretical description to within
statistical errors of simulation and the numerical
approximations used.
\item[(iii)] The results of simulation match the results of
experimental studies of the critical behavior of diluted
Ising-like magnets.
\end{description}

\begin{acknowledgments}
The authors thank D.P.~Landau, M.~Novotny,
V.~Yanke, and N.~Ito for fruitful discussions of the
results of this study during the 3rd International Seminar
on Computational Physics in Hangzhow (China,
November 2006).

This study was supported by the Russian Foundation
for Basic Research (project nos. 04-02-17524 and
04-02-39000) and partly by the Program of the President
of the Russian Federation (grant no. MK-8738.2006.2).

The authors are grateful to the Interdepartmental
Supercomputer Center of the Russian Academy of Sciences
for providing computational resources.
\end{acknowledgments}

\end{document}